\documentclass[fleqn,usenatbib]{mnras}
\usepackage{newtxtext,newtxmath, mathtools}
\usepackage[T1]{fontenc}
\usepackage{ae,aecompl}
\usepackage{graphicx}	% Including figure files
\usepackage{amsmath}	% Advanced maths commands
\usepackage{amssymb}	% Extra maths symbols
\usepackage[dvipsnames]{xcolor}
\usepackage{natbib}
\usepackage{subfig}
\usepackage{ mathrsfs }
\usepackage{epstopdf}
\usepackage{tipa}
\usepackage{ulem}

\newcommand{\Sdirhatnu}{{S_{\rm dir,\nu}}}
\newcommand{\tauhatnu}{{\hat \tau_\nu}} 
\newcommand{\tauhatmax}{{\hat \tau_{\nu,{\rm max} }}} 
\newcommand{\Fscatnu}{F_{{\rm sc},\nu}}
\newcommand{\Lscatnu}{L_{{\rm sc},\nu}}
\newcommand{\Forcescatnu}{\mathscr{F}_{{\rm sc},\nu}}
\newcommand{\phiscnu}{\Phi_{{\rm sc},\nu}}
\newcommand{\phidirnu}{\Phi_{{\rm dir},\nu}}
\newcommand{\phiIRthin}{\Phi_{{\rm IR,thin}}}
\newcommand{\phithin}{\Phi_{{\rm thin}}}
\newcommand{\rbarscatnu}{ \left< r \right>_{\mathscr{F},{\rm sc},\nu }}
\newcommand{\Tdin}{{T_{d,{\rm in}}}}

\newcommand{\rin}{{r_{\rm in}}}
\newcommand{\rout}{r_{\rm out}}
\newcommand{\rhoin}{{\rho_{\rm in}}}
\newcommand{\rph}{{r_{\rm ph}}}
\newcommand{\tauinfnu}{\tau_{\infty,\nu}} 
\newcommand{\force}{{{F}_{\rm rad}}}
\newcommand{\taufid}{\tau_{\rm fid}}
\newcommand{\forcenu}{{ F}_{\rm rad,\nu}}
\newcommand{\forcetot}{{ F}_{\rm rad, tot}}
\newcommand{\rF}{\left< r \right>_{{ F}}}
\newcommand{\rFthin}{\left< r \right>_{{F,{\rm thin} }}}
\newcommand{\rFthick}{\left< r \right>_{{F,{\rm thick} }}}

\newcommand{\Lstarnu}{{L_{*,\nu}}}
\newcommand{\Ldirnu}{{L_{\rm dir,\nu}}}
\newcommand{\Phidir}{{\Phi_{\rm dir}}}
\newcommand{\Phisc}{{\Phi_{\rm sc}}}
\newcommand{\PhiIR}{{\Phi_{\rm IR}}}
\newcommand{\PhiIRthick}{{\Phi_{\rm IR,thick}}}
\newcommand{\HII}{H\;{\scriptsize II}\ }
\newcommand{\Prad}{{P_{\rm rad}}}

\newcommand{\cdm}{ } %%%\color{black}
\newcommand{\cdmnew}{ } %\color{purple}
\newcommand{\phjedit}{ } % \color{black}
\newcommand{\phjaddition}{ }  %%%\color{blue}
\newcommand{\phjedittext}{ } %%%\color{teal}
\title[ Radiation Forces on Dust Envelopes ]{ Radiation Forces on Dust Envelopes } 
\author[Jumper \& Matzner]{
Peter H. Jumper,%$^{1}$%\thanks{E-mail: mn@ras.org.uk (KTS)}
\ Christopher. D. Matzner%,$^{2}$
\\
$^{1}$Department of Astronomy and Astrophysics, University of Toronto \\ 
}

\date{Accepted XXX. Received YYY; in original form ZZZ}

\pubyear{2017}

\begin{document}
\label{firstpage}
\pagerange{\pageref{firstpage}--\pageref{lastpage}}
\maketitle

%%%%%%%%%%%%%%%%%%%%%%%%%%%%%%%%%%%%%%%%%%%%%%%%%%%%%%%%%%%%%%%%%%%%%%%%%%%%%%%%%%%
%%%%%%%%%%%%%%%%%%%%%%%%%%%% ABSTRACT OF THE PAPER %%%%%%%%%%%%%%%%%%%%%%%%%%%%%%%%
%%%%%%%%%%%%%%%%%%%%%%%%%%%%%%%%%%%%%%%%%%%%%%%%%%%%%%%%%%%%%%%%%%%%%%%%%%%%%%%%%%%
\begin{abstract} 

{{We address in detail the radiation forces on spherical dust envelopes around luminous stars, and numerical solutions for these forces, as a first step toward more general dust geometries. }} 
Two physical quantities, a normalized force and a force-averaged radius,  suffice to capture the overall effects of radiation forces; {\cdmnew these combine to yield the radiation term in the virial theorem}. In addition to the optically thin and thick regimes, the wavelength dependence of dust opacity allows for an intermediate case in which starlight is easily trapped but infrared radiation readily escapes.  Scattering adds a non-negligible force in this intermediate regime.  We address all three regimes analytically and provide approximate formulae for the force parameters, for arbitrary optical depth and inner dust temperature.  Turning to numerical codes, we examine the convergence properties of the Monte Carlo code Hyperion run in Cartesian geometry.  We calibrate both Hyperion and our analytical estimate using the DUSTY code, run in spherical geometry. We find that Monte Carlo codes tend to underestimate the radiation force when the mean free path of starlight is not well resolved, as this causes the inner dust temperature, and therefore the inner Rosseland opacity, to be too low. We briefly discuss implications for more complicated radiation transfer problems. 

\end{abstract}
 
% Select between one and six entries from the list of approved keywords.% Don't make up new ones.

\section{Introduction}  

Dusty gas is much more opaque to visible light than ionized gas lacking dust.  As a result, radiation pressure forces  become dominant in situations where luminous but sub-Eddington objects, like massive stars and AGN, are surrounded by dusty gas.  Examples include
individual massive star formation, where radiation forces present an obstacle to stellar accretion \citep{wolfire87}; massive star cluster formation \citep{2010ApJ...710L.142F,matznerjumper15}, in which matter may be expelled from the cluster-forming zone; the disruption of giant molecular clouds  \citep{km09,2010ApJ...709..191M,hopkinsetal12}; the initial inflation of giant \HII regions \citep{draine11,2011ApJ...731...91L,2012ApJ...757..108Y};\footnote{Dust radiation transfer is more important in luminous \HII regions than one might expect, because grain absorption exceeds ionization absorption precisely when radiation forces are strong; see \citet{2012ApJ...757..108Y}. } and the ejection of gas from galaxies \citep{mt10}; as well as dust-driven winds and superwinds from AGB stars \citep{1991ApJ...375L..53B}.   

For each of these problems a detailed understanding of radiation forces on dust, and a calibration of numerical methods to estimate these forces, are clearly required.  
%%%%%{\phjedittext especially as instabilities, such as the Rayleigh-Taylor instability,  may have a profound effect on the emergent solution.  In simulations, the growth and saturation of radiation-driven instabilities are sensitive to the radiative transfer method. For example, simulations have reached opposite conclusions regarding the stability of super-Eddington star formation within galaxy disks \citep{2013MNRAS.434.2329K,2014ApJ...796..107D} and the of radiation-dominated accretion disks \citep{2009ApJ...691...16H,2014ApJ...784..169J}, depending on whether radiation was treated with flux-limited diffusion or a variable Eddington tensor technique.  Such closure techniques are fast enough to be implemented in dynamical simulations, but incur an error that must be calibrated and understood. 
%%%%%}
As a step toward a more complete understanding, we focus here on a dramatically simplified case:  a spherically symmetric, power-law dust profile surrounding a central light source.   
{\cdmnew We choose this problem several goals in mind.   First, we wish to establish a set of measures (output parameters) with which the effects of radiation forces can be summarized.  Second, we aim to understand in analytical terms how these measures depend on the innermost dust temperature and optical depth of the envelope (input parameters).     

Lastly, we wish to calibrate Monte Carlo (MC) calculations of finite resolution: as we shall show,  MC results suffer a type of systematic error when the photon mean free path is not well resolved.  Understanding errors in the MC technique will be useful for our future work, in which we use moderate-resolution MC simulations to examine radiation forces in inhomogeneous dust distributions that break spherical symmetry, and then characterize errors in approximate techniques.  More generally, resolution-dependent errors are dynamically important within simulations; for instance, in the onset of Rayleigh-Taylor instabilities during massive star formation (see \citealt{2011ApJ...732...20K} and \citealt{2016MNRAS.463.2553R}). 
}

%show that radiation forces in a simple spherical problem can be understood in analytical terms. Second, we aim to characterize systematic errors that persist even in Monte Carlo solutions when the dust density must (e.g., because of the constraints imposed by dynamical codes) be represented at coarse resolution.  Third, we wish to set the stage for an upcoming generalization to clumpy envelopes in which the spherical idealization is relaxed. Our larger goal is to calibrate Monte Carlo as a tool for evaluating moment-closure techniques not considered in this paper, and we are motivated by the fact that several instabilities and their saturation have been observed to depend on the radiation transfer method.

%{\phjcollab (PROPOSED: TAKE THE MATERIAL RELEVANT TO THE CITATIONS AND BRING IT DOWN HERE.  THEN WE NEED TO BRDIDGE THE GAP.  WE CAN ALSO BRING UP RESOLUTION EFFECTS WITH \citet{2011ApJ...732...20K} and \citet{2016MNRAS.463.2553R} ) }  

%%%% cdm - this paragraph is no longer necessary. %%%%

Our spherical problem can be approached by the multi-group radiation transfer code DUSTY \citep{dustyuser}{\phjaddition,} 
%{ \phjdeletion \sout{ 
whose adaptive spatial and frequency grids allow it to rapidly converge  to a solution that we shall consider to be ground truth.  
But, DUSTY cannot treat complicated, three dimensional dust distributions.  {\cdmnew For these we employ the MC code Hyperion \citep{robitaille11}, modified to record radiation forces.  In our numerical sections we focus on the convergence of radiation force measures, calculated within Hyperion, toward a value determined by DUSTY.  This is not meant to be a code comparison (for which it would be more appropriate to run Hyperion in spherical symmetry). }

%Monte Carlo methods should also converge to ground truth. However Monte Carlo convergence is rather different from DUSTY's, { as it relies on sufficiently high spatial} and spectral resolution,
%as well as the propagation of sufficiently many photon packets. 
%} } 
%
%{\phjdeletion \sout{
%Furthermore, the weaknesses of moment-closure methods will be most apparent in non-spherical problems in which the radiation pressure tensor is anisotropic. 
%} }

%For these reasons we shall compare DUSTY to 

On the analytical front, radiative transfer through spherical dust envelopes, and the corresponding dust temperature profiles, have long been studied in relation to the emergent spectral energy distributions of protostars \citep{Larson69,adamsshu85}, star cluster-forming clumps in starburst galaxies  \citep{cm05}, and dusty winds from late-type stars \citep{ivezicelitzur95}.  
%Note that the radiative transfer solutions of a spherical shell can be described in terms of scaling relations with respect to normalized radial profiles,a point made by  \citealt{RowanRobinson80} and leveraged in later solutions and in the DUSTY code.
%%%%%{\phjcommentnew - Insert here a statement about other FLD methods - Krumholz's involvement with HARM2, FLD, and some of Peters results.}
%%
%%Hybrid Adaptive Ray-Moment Method (HARM2): A Highly Parallel Method for Radiation Hydrodynamics on Adaptive Grids
%By  Anna L. Rosen,Mark R. Krumholz,Jeffrey S. Oishi,Aaron T. Lee,Richard I. Klein
%http://adsabs.harvard.edu/cgi-bin/bib_query?arXiv:1607.01802
%%
%%The reliability of approximate radiation transport methods for irradiated disk studies
%Rolf Kuiper,Ralf S. Klessen
%%We check the accuracy of the gray flux-limited diffusion (FLD) approximation and a gray and frequency-dependent ray-tracing plus FLD approximation.
%%https://arxiv.org/abs/1305.2197
%%http://adsabs.harvard.edu/cgi-bin/bib_query?arXiv:1305.2197
%%
%%
%% THOMAS PETERS – Arxiv 
%https://arxiv.org/find/astro-ph/1/au:+Peters_T/0/1/0/all/0/1
%Selected papers –
%Radiative transfer calculations of the diffuse ionised gas in disc galaxies with cosmic ray feedback
%The turbulent life of dust grains in the supernova-driven, multi-phase interstellar medium
%Simulating the Formation of Massive Protostars: I. Radiative Feedback and Accretion Disks
%Etc. 
%%
%%
However, analytical studies (including \citealt{km09}, \citealt{2010ApJ...709..191M}, and our own work: \citealt{matznerjumper15}) have so far considered only crude approximations to the radiation force.  We will develop more accurate analytical formulae, albeit restricted (for now) to the simple spherical geometry.

%%%%%%%%%%%%%%%%%%%%%%%%%%%%%%%

%{\phjcommentnew Krumholz, Kuiper, Peters - Radiation Pressure of dusty envelopes}

%%%%%%%%%{\phjaddition We also consider the }
 
We delineate the problem to be solved below in \S \ref{SS:ProblemSetup}.  In \S \ref{S:Quantities_of_interest} we introduce two integral quantities of interest, the net radiation force and the force-averaged radius, which can be combined to form a radiation force term in the virial equation.  We address the theoretical problem in \S \ref{S:Analytics} and compare numerical solutions in \S \ref{S:CodeComparison}.  

\subsection{Physical problem} \label{SS:ProblemSetup}
We consider, for simplicity, a spherically symmetric dust envelope with inner radius $\rin$, outer radius $\rout$ (set to $4\rin$ in our fiducial case, purely for convenience) and radial profile $\rho(r)\propto r^{-k}$ for $\rin<r<\rout$, for some $k>1$;  in our fiducial case $k=1.5$.  
This is representative of the density profile in low and high-mass star formation:  \citet{vandertak00} finds $k=1.0$ to $k=1.5$,  \citet{Jorgensen02} finds $k = 1.3$ to $k=1.9$, $\pm 0.2$, \citet{shirley02} finds $k = 1.8 \pm 0.1$, and \citet{Mueller02} finds $k = 0.75$ to $k = 2.5$ with a mean value of $\langle k \rangle = 1.8 \pm 0.4$. 

 We  take the spectrum of the central source, $\Lstarnu$, to correspond to a blackbody of color temperature $T_*=5772$\,K.  We adopt the dust absorption opacity $\kappa_{a,\nu}$ and albedo $a_\nu$ for a dust mixture with $R_V=5.5$ provided by  \citet{draine03,draine2003b} %\citet{weingartnerdraine01}, 
from which we compute the total opacity {$\kappa_\nu=\kappa_{a,\nu}(1+a_\nu)$.}  
However, to treat scattered radiation identically in numerical and analytical calculations, we consider only isotropic scattering.
{ This neglects the characteristic mean scattering angle, $\langle cos\left( \theta \right) \rangle$, included in the  \citet{draine03,draine2003b} dust model, and thus does not consider the effects of realistic phase functions, which may have a preferential direction of scattering.}
 %% REV 1 VERSION {This neglects a scattering angle $\theta $ characterized by a mean angle of $\langle cos\left( \theta \right) \rangle$. }
 %{\cdm  This requires that we adjust $a_\nu$ to a value that achieves the same momentum transfer in the isotropic cases.} 
While this is certain to alter our results to a small degree, it also permits a precise comparison between various approaches. 

\subsection{Quantities of Interest}\label{S:Quantities_of_interest}
%%% UPGRADE \mathcal{R} references.... 

%{\phjeditcom \noindent SHALL WE PROMOTE $\mathcal{R}$?}

We take the outward luminosity at radius $r$ to be $L(r) = \int_0^\infty L_\nu(r)\, d\nu$; photon momentum passes $r$ at the rate $L(r)/c$, which is independent of $r$ in static equilibrium (whereas $L_\nu$ can vary with radius). 
Given an extinction optical depth $\tau_\nu$ (arising from a density and specific opacity, $d\tau_\nu = \rho(r)\,\kappa_\nu\,dr$), the radiation force satisfies
\begin{equation} \label{eq:forceGeneral} 
d\forcenu = \frac{L_\nu}{c} d\tau_\nu;
\end{equation}
the frequency-integrated force within $r$ is $\force(r) = \int_0^\infty \forcenu\, d\nu$, and the total outward force is $\forcetot = \force(r=\infty)$. 

Our first quantity of interest, then, compares the applied radiation force to the photon force: 
\begin{equation}\label{eq:def-Phi} 
\Phi \equiv {\forcetot \over L/c} = \int_0^\infty  \frac{L_\nu}{L}\, d\tau_\nu\,d\nu. 
\end{equation} 
This can be less than unity, in the case of an optically thin dust envelope, or much greater than unity if the optical depth is very high.   Indeed $\Phi$ is a luminosity-weighted integral of $\tau_\nu$, as the final expression in equation (\ref{eq:def-Phi}) shows.   

It is also important to know where the force is applied.  For this we introduce a second quantity, the force-averaged radius:
\begin{equation}\label{eq:def-<r>_F} 
\rF \equiv \int_0^{\forcetot} r\, d\force /\forcetot. 
\end{equation}  
 We are motivated here by the virial theorem, where the radiation force ${\mathbf F}_{\rm rad}$ introduces the term ${\cal R} = \int {\mathbf r}\cdot d{\mathbf F}_{\rm rad}$,  equivalent to expression (4.24) of \citet{mz92}.  For our spherically symmetric problem,
 \begin{equation}\label{virial-R} {\cal R} =  \Phi \frac{L}{c} \rF.\end{equation}

Our model density distributions are described by inner and outer radii $\rin$ and $\rout$.  Dimensionless quantities like $\Phi$ and $\rF/\rin$ are functions of the spectral shape of the input radiation (or its color temperature, if it is taken to be a blackbody) and the optical depth $\tau_{\rm fid}$ at a chosen reference frequency $\nu_{\rm fid}$ (see \citealt{rowanrobinson80} and \citealt{ivezicelitzur97}).    

For brevity we use $\tau_\nu(r)$ to denote the dust extinction optical depth within $r$, and $\tau_\nu$ to denote the total optical depth through the dust distribution.  Therefore $\tau_\nu = \tau_\nu(\infty)$. 

\section{Analytical Predictions  }\label{S:Analytics} 

Dusty radiative transfer has two quintessential features: first, the dust opacity $\kappa_\nu$ increases with frequency over the relevant range of $\nu$; and second, the stellar surface temperature is much hotter than dust grains can possibly be.   Therefore, optical and ultraviolet starlight is always more readily absorbed than the infrared emission from heated grains.
 As a result, one can distinguish three regimes: {\bf I} -- optically thin to starlight ($\tau_*<1$ where $\tau_*= \int_0^\infty \tau_\nu (L_{*,\nu}/L)\,d\nu$ is the starlight-averaged opacity); {\bf II} -- thick to starlight but thin to dust emission ($\tau_* > 1 > \tau_R(\Tdin)$ where $\tau_R(\Tdin)$ is the Rosseland opacity at the temperature of the innermost dust ); and {\bf III} -- optically thick to dust radiation ($\tau_* > \tau_R(\Tdin) > 1$).

\noindent{\bf I. Optically thin case: Ia.\ Starlight. } Of these, the optically thin case is of course the simplest.   The direct starlight luminosity at $r$ is $\Ldirnu(r)=e^{-\tau_\nu(r)} \Lstarnu$, and therefore the contribution of direct irradiation to $\Phi$ is \footnote{Of course $\Phidir=\tau_*+{\cal O}(\tau_*^2)$ would be equally valid, but expression (\ref{eq:Phi_direct}) is more accurate when $\kappa_\nu$ varies slowly with $\nu$.}
\begin{equation} \label{eq:Phi_direct}
\Phidir = \int_0^\infty (1-e^{-\tau_\nu}) {L_{*,\nu}\over L}\, d\nu = 1 - e^{-\tau_*} + {\cal O}(\tau_*^2).
\end{equation}

In the limit $\tau_*\ll 1$ of very low optical depth, $e^{-\tau_\nu}\rightarrow 1$ and $\rF$ approaches a unique value $\rFthin$; for our truncated power law profile this is  
\begin{equation}\label{eq:rF_direct_OptThin}
\rFthin \equiv {k-1\over 2-k} \rout^{2-k}\rin^{k-1} {1-(\rin/\rout)^{2-k} \over 1-(\rin/\rout)^{k-1} },
\end{equation} 
which is unity for a thin shell ($\rout =\rin$) and is intermediate between $\rin$ and $\rout$ even for very thick shells, so long as $1<k<2$.  In our fiducial case $k=1.5$, $\rFthin = (\rin \rout)^{1/2}$. 

\noindent{\bf Ib. Optically thin dust emission.} 
A minor but non-negligible contribution to the total force arises from the interaction of thermal dust radiation with other dust grains.  The total infrared luminosity is approximately $(1-e^{-\tau_*})L$ and the appropriate opacity is $\kappa_{dd}(\Tdin)$, where 
\begin{equation}\label{eq:kappa_dd}
\kappa_{dd}(T) = {\int_0^\infty B_\nu(T) \kappa_\nu \kappa_{a,\nu}\,d \nu  \over \int_0^\infty B_\nu(T) \kappa_{a,\nu}\,d \nu} 
\end{equation}
is the opacity of dust grains to the emission of other grains at temperature $T$. The associated optical depth is $\tau_{dd,{\rm in}} = [\kappa_{dd}(\Tdin)/\kappa_{\rm fid}] \taufid$.  Note that $\kappa_{dd}(T)$ is relatively large compared to the Rosseland mean, because optically thin radiation is concentrated in frequencies where $\kappa_\nu$ is maximized; otherwise optically thin infrared would be entirely negligible. 

The contribution to $\Phi$ due to recaptured infrared emission is therefore approximately
\begin{equation}\label{eq:PhiIRthin}
\phiIRthin = { { \kappa_{\rm a,*} \over \kappa_*} } (1-e^{-\tau_*})(1-e^{-\tau_{dd,{\rm in}}})
\end{equation}
and we can estimate  the total force due to optically thin radiation as \[\phithin = \Phi_{\rm dir} + \phiIRthin.\]  

\noindent{\bf II. Intermediate case:  starlight scattering and absorption. }  As the starlight optical depth increases, $\Phidir\rightarrow 1$ but scattered starlight adds to the net force and hence a new component $\Phisc$ to the force ratio $\Phi$.  We address this case in Appendix \ref{S:ScatteredLightAppendix} by means of Eddington's approximation.  In the limit $\tau_*\gg1$ (\S\,\ref{SS:HighTauScattering}), $\Phidir\rightarrow 1$ and 
\begin{equation}
\Phisc \rightarrow \left< {a_\nu \over 1 + \sqrt{3(1-a_\nu)}}  \right>_{L_*} \simeq  {a_* \over 1 + \sqrt{3(1-a_*)}} 
\end{equation} 
where $\left<\cdots\right>_{L_*}$ is a starlight-averaged value, i.e., a frequency average weighted by $L_{*,\nu}$, and $a_* = \left<a_\nu\right>_{L_*}$.    

In the asymptotic case $\tau_R(\Tdin)\ll 1 \ll \tau_*$, all the force is applied at the inner boundary and therefore $\rF/\rin\rightarrow 1$.  Figure \ref{fig:EddingtonScattering} demonstrates the dependence of $\Phidir + \Phisc$ and the force-averaged radius due to (direct + scattered) starlight, as functions of optical depth and albedo. 

\noindent{\bf III. Optically thick case. } Here the dust is optically thick to starlight, and also to its own infrared thermal radiation.   The direct and scattered radiation and their associated moments { (like $\Phidir$ and $\Phisc$) }  are as described above, but the self-force due to dust emission becomes appreciable and contributes a new term $\PhiIR$. In Appendix \ref{S:Diffusion} we compute $\PhiIR$ using the diffusion approximation, deriving the thick limit
\begin{eqnarray}  \label{eq:PhiIR}
\PhiIRthick &\simeq& {4-\beta \over 4(k-1)+2\beta } \kappa_{R, {\rm in}} \rho_{\rm in} \rin  \nonumber 
\\&=& {(4-\beta)(k-1) \over 4(k-1)+2\beta } { \kappa_{R, {\rm in}} \over \kappa_{\rm fid} } 
{\tau_{\rm fid}\over  1-\left(\rin/\rout \right)^{k-1}}.
\end{eqnarray} 

The latter expression (in terms of the inner conditions as well as a fiducial opacity and optical depth) is appropriate for comparison to DUSTY simulations.  It demonstrates that the normalized radiation force depends on the Rosseland opacity at the inner boundary, and is therefore directly related to the inner dust temperature; this point will be relevant to our analysis of numerical methods.  

In the thermal diffusion limit $\rF$ takes a limit $\rFthick$ given in equation (\ref{eq:rF_diffusive}) of Appendix (\ref{S:Diffusion}). 

\noindent{\bf Combined formulae.} To combine these asymptotic forms, we propose 
\begin{equation}\label{eq:combined-Phi}
\Phi \simeq \left(1-e^{-\tau_*}\right)\left[1 + { { \kappa_{\rm a,*} \over \kappa_*} }(1-e^{-\tau_{dd,{\rm in}}})  + \Phisc \left(1-e^{-\tau_*}\right) + \PhiIRthick \right]
\end{equation} 
and {
\begin{equation}\label{eq:combined-rF} 
\rF \simeq  \left[\rin +\left(\rFthin-\rin\right) e^{-{\tau_*\over q}}   \right] \left(1 - {\PhiIR'\over \Phi}\right) + \rFthick {\PhiIR'\over \Phi},
\end{equation} 
where $\PhiIR' = (1-e^{-\tau_*})\PhiIRthick$. The radiation contribution $\cal R$ in the virial theorem is then given by 
\begin{equation}\label{eq:combined-calR} 
{{\cal R}\over L/c} \simeq  \left[\rin +\left(\rFthin-\rin\right) e^{-{\tau_*\over q}}   \right] \left(\Phi-\PhiIR'\right) + \rFthick \PhiIR'.
\end{equation} 

Note that we include a prefactor $1-e^{-\tau_*}$ on $\PhiIRthick$ in equation (\ref{eq:combined-Phi}), since IR luminosity is powered by absorbed starlight.  
The parameter $q$ in equation (\ref{eq:combined-rF}) and (\ref{eq:combined-calR}) does not affect the asymptotes, so it is somewhat arbitrary; we find that $q\simeq10$ optimizes the fit.  
}

\section{\cdmnew Radiation transfer codes} \label{S:CodeComparison}

{\cdmnew 
We now turn to the numerical codes DUSTY and Hyperion.  We use fully converged, variable resolution DUSTY simulations to provide precise values of the radiation force measures $\Phi$, $\rF$, and $\cal R$ across our parameter survey, which we use to check the accuracy of our analytical predictions.  Our Hyperion runs serve to explore the convergence properties of moderately resolved, three-dimensional Monte Carlo simulations.  This calibration will be useful for numerical hydrodynamics simulations, and also for our own future work on radiation forces in inhomogeneous dust distributions. 
}

%to examine their convergence with respect to our quantities of interest, $\Phi$ and $\rF$ and to check the accuracy of our analytical predictions.  {\phjedittext We are also particularly interested in the errors accrued in Monte Carlo methods, like Hyperion, as checked against the solutions of the other methods.  }  
%%%%{\phjedittext} 
 
\subsection{DUSTY: adaptive radiation transfer code}\label{SS:Dusty}

\citet{ivezicelitzur97} found that the radiative transfer equations, if given a specified temperature for the inner edge of the dust envelope and the SED of the luminosity source, could be solved without further reference to dimensional values; everything else % the rest of the parameters
could be expressed 
%and utilized 
in terms of dimensionless quantities varying with respect to a dimensionless radial profile.  %A similar set of principles are utilized in the DUSTY code.  
Using these methods, DUSTY \citep{dustyuser} 
provides solutions to the radiative transfer problem parameterized by the inner dust temperature $\Tdin$, {\phjedittext the ratio $n$ relating the outer and inner radii such that $r_{\rm out} = n r_{\rm in} $, the density power law index $k$}, the optical depth $\taufid$ at a fiducial frequency $\nu_{\rm fid}$, and the spectral distributions of the central source ($\Lstarnu$)  and of the dust properties ($\kappa_\nu$, $a_\nu$).   We inspect these solutions to construct $\Phi$ and $\rF$.  

%{\phjedittext To calculate these solutions, DUSTY creates a self-refined computational grid along the radial profile.  
%The initial grid is constructed such that the following quantities change by no more than a specified value between consecutive grid points:  the increase in the ratio $r/r_{\rm in}$, the ratio of the densities at these points, and the increment in the fraction of the total fiducial optical depth.  For these values, we take 2.0, 4.0, and 0.025 (default: 0.3) to start.  DUSTY then refines this grid in an iterative process by calculating the conservation of bolometric fluxes throughout and then decreasing the step size (and thus also adding more grid points) if the flux is not conserved within a specified accuracy. 
%%% The userdefined parameter npY imposes a limit on the number of grid points allowed.  
%In our survey, we start with a requested flux accuracy of 0.01 (default: 0.05) }

{\phjedittext To calculate these solutions, DUSTY creates a self-refined grid along the radial profile through an iterative process.  Starting at the inner cavity wall, $r/r_{\rm in} =1$, DUSTY creates an initial set of radial grid points such that for each pair of consecutive points, each of three regulating quantities changes by less than some specified increment or ratio.  These quantities are the increment of the optical depth to the fiducial frequency as a fraction of the total depth $\taufid$ (default: 0.3), the ratio of their radii (default: 2.0), and the ratio of their densities (default: 4.0).  We preserve the latter two of these defaults, while adopting a different value for the maximum fractional increment of the optical depth, $0.025$, as a starting value which we will soon vary.  Once the initial grid has been generated, DUSTY begins the self-refinement by calculating the bolometric flux through the grid, checking whether the flux is conserved and converged within some specified accuracy parameter (default: 0.05).  If it is not, or if too large a fraction of the allowed accumulated error is accrued between a single pair of points, DUSTY adds additional grid points in the vicinity of these problematic points to further refine the problem, iterating this procedure until an accepted grid is produced.  DUSTY calculates the radiative transfer solutions by determining the spectral energy density with an integral equation \citep{ivezicelitzur97}, utilizing a temperature profile derived from the assumption of radiative equilibrium. 

Adopting a starting flux accuracy parameter of 0.01 for a quick parameter survey of DUSTY's convergence, we vary both this value and the maximum fractional increment of the optical depth to explore this space.  For these, we hold $\Tdin$ fixed at 1500\,K (roughly the dust sublimation temperature) and set $\taufid=10$, which for $c/\nu_{\rm fid}=1.95\,\mu$m corresponds to $\tau_*= 54.7$. The results are shown in Tables \ref{tab:fluxcon} and \ref{tab:delTAUsc}.    DUSTY's solutions { appear to converge as the requested tolerances are made more strict and more grid points are required to determine the solution. {  Fitting the results with linear functions of the tolerances, we infer  $\Phi\rightarrow  5.183$ and $\rF\rightarrow 1.596\,\rin$ as these tend to zero. }  } 
}

\begin{table} 
\begin{centering}
\caption{Convergence of DUSTY radiative transfer for the problem with $\Tdin= 1500$\,K and $\taufid=10$ at $c/\nu_{\rm fid}=1.95\,\mu$m.  Here we vary the requested flux conservation accuracy and holding all other control parameters and physical parameters of the problem constant.  For requested accuracy of 0.01 or worse, other control parameters provide more stringent refinement criteria and the solution is unchanged.  At higher requested accuracy, more grid points are finally required, and  $\Phi$ varies slightly (a 1\% change as the requested accuracy changes over an order of magnitude).{ \phjedit Meanwhile, $\langle r \rangle_F/r_{\rm in}$ remains  insensitive to the required flux accuracy }}    
 \label{tab:fluxcon}
\begin{tabular}{ |c |c |c |c |c |}
 \hline {Req.\ Accuracy} &  {Grid Points} & {$\Phi$} &  {$\langle r \rangle_F/r_{\rm in}$} & Flux  Accuracy \\
\hline 0.001 & 73 & 5.20 & 1.59 & $3.0\times10^{-4} $ \\ %{ Re-running the model [couldn't locate the original directory], get a slightly different Grid Point value, 72 -> 73}
\hline 0.005 & 60 & 5.22& 1.59 & $1.7\times10^{-3} $ \\
\hline 0.01 & 57 & 5.25& 1.59 &  $3.6\times10^{-3}$ \\
\hline 0.05 & 57 & 5.25& 1.59 & $3.6\times10^{-3 }$ \\
\hline 0.10 & 57 & 5.25 & 1.59 & $3.6\times10^{-3}$ \\
\end{tabular}
\end{centering}
\end{table}

\begin{table} 
\begin{centering}
\caption{DUSTY convergence for the same problem as Table \ref{tab:fluxcon}, here varying the maximum step in optical depth between consecutive grid points relative to the total optical depth, $\max \Delta \tau/\tau$, whilst holding all other control variables at default values. Allowing 8 times larger steps, number of steps decreases by a factor of 4.5;  $\Phi$ changes by about 1.3\% from its starting value, {\phjedit while $\langle r \rangle_F/r_{\rm in}$ chances by about 0.63\% from its starting value. } }\label{tab:delTAUsc}
\begin{tabular}{ |c |c |c |c |c |}
 \hline {\rm Max  $\Delta \tau / \tau$} &  Grid Points %{\em nY}
& {$\Phi$} &{$\langle r \rangle_F/r_{\rm in}$} &  Flux Accuracy \\
\hline 0.0125 & 108 & 5.23 &1.59  & 3.7$\times 10^{-3}$\\
\hline 0.025 & 57 & 5.25& 1.59 & 3.6$\times 10^{-3}$ \\
\hline 0.05 & 33 & 5.29& 1.59 & 3.3$\times 10^{-3}$\\
\hline 0.1 & 24 & 5.32&  1.58 & 4.7$\times 10^{-3}$ \\
\hline 0.2 & 23 & 5.32& 1.58 & 6.6$\times 10^{-3}$ \\
\end{tabular}
\end{centering}
\end{table}

\subsection{Hyperion: Monte Carlo radiation transfer code} \label{SS:Hyperion}

 Hyperion \citep{robitaille11}  represents the alternative Monte Carlo technique. Monte Carlo codes  like Hyperion model radiation transfer by propagating a large number of photon packets throughout the medium,  allowing them to  scatter or be absorbed and re-emitted in a number of interactions. 
 
 {\phjedittext Hyperion has the advantage of being able to handle arbitrary density distributions with a choice of several coordinate geometries.  Thus, it can solve more types of problems than can DUSTY, which is limited to spherically symmetric or slab geometries. We adopt a Cartesian coordinate geometry here to maintain consistency with a subsequent paper, in which we will generate inhomogeneities as clumps in the dust envelopes; this process is simplified in such a geometry.  However, Hyperion is far slower than DUSTY for the spherically symmetric problem, which they can both solve; and while it can accept hierarchically refined density distributions, it cannot adaptively re-grid. }
%%%% EDIT ---> This is actually another good point pertaining to the forces... 
  
{\phjedittext In its base form, Hyperion calculates the specific energy absorption rate in the medium due to the radiative transfer.  This allows one to derive properties such as temperatures, the spectral energy distribution, and resulting images, but disregards and discards vector information.  Therefore, to calculate the radiation forces exerted by the photons, we modified Hyperion to record and track the specific force (an acceleration, the force per unit mass) exerted upon the material in each cell as a vector with Cartesian components.  The specific force in each cell is built up as a summation of the contributions by two mechanisms from each photon packet: their interaction events with the dust, and their propagation through the dust. The former includes all scattering and absorption events, which are treated as momentum transfer events based on the change in the packet's energy and velocity vector.  The latter reflects the weighted Monte Carlo technique of  \citet{1999A&A...344..282L},
%%%%{\phjeditcom Lucy 1999}
which deposits force along the entire path of propagation.  {\cdmnew Our approach is similar to, but independent of, the work by \citet{harries15}. }
%%%% THIS COMMENTED VERSION COMMENTS OUT THE HARRIES COMMENTS - SHOULD WE PERHAPS ADD IT... 
%%%%%We  modified Hyperion to record the vector field of radiation forces associated with photon absorption and scattering; our approach is similar to, but independent of, the work by \citet{harries15}. 
}

Whereas DUSTY's simulations are parameterized by $\Tdin$ and $\taufid$, Hyperion's are parameterized by dimensional quantities like distance, luminosity, and density (as defined on the computational grid).  {\phjedittext The dust emissivity and mean opacities are tabulated as a function of the specific energy absorption rates based on the absorption and emission of energy, and the temperatures may be calculated in turn as a function of these rates, converging toward the ideal solution just as DUSTY does.}
%%% CORRECTION
%Dust temperatures are determined self-consistently through the absorption and emission of energy, and these converge toward the ideal solution just as DUSTY does. 
{\phjedittext We choose $L_*=L_\odot$ ({\cdmnew an arbitrary choice that can be adjusted without loss of generality}), adopt the same stellar spectrum and dust properties, and then set the dust density coefficient and $\rin$ (here $2.15\times 10^{12}$\,cm in the case of $\tau_{\rm fid} = 10$) so that the converged values of $\Tdin$ and $\taufid$ match the DUSTY calculation.}
%We choose $L_*=L_\odot$, adopt the same stellar spectrum and dust properties, and then set the dust density coefficient and $\rin$ (here $2.15\times 10^{12}$\,cm) so that the converged values of $\Tdin$ and $\taufid$ should match the DUSTY calculation. 

Hyperion's results must converge in two ways.  First, there must be sufficiently many photon packets propagated through the grid for the dust temperature distribution to settle toward equilibrium.  Although small-number statistics certainly add variance to the dust temperatures, we found that the recommended 30 packets per grid cell was easily sufficient (Tables \ref{tab:photoncell} and \ref{tab:packetcell}).  Indeed, integral quantities like $\Phi$ and $\rF$ appear to be remarkably insensitive to discrete sampling of the photon field.

\begin{table} 
\begin{centering}
\caption{Variation of Hyperion's predictions for $\Phi$ as the numbers of photon packets per cell are varied, in the same problem as before. Cases labeled ``Low'', ``Medium'', and ``High'' correspond to averages of 10, $10^6/32^3=30.5$, and 100 photon packets per cell, respectively.  } \label{tab:photoncell}
\begin{tabular}{ |c |c |c |c |}
 \hline% 
 {\# Cells} &% 
 %&  {10 phot./cell } & {$10^6/32^3$ phot./cell  } &  {100 phot./cell }\\
 %&
 Low: $\Phi$ & Medium: $\Phi$ &High: $\Phi$ \\
\hline $32^3$& 4.735 & 4.734 &  4.734 \\ %%%4.73462 & 4.73383 &  4.73398 \\
\hline $64^3$ & 4.932  & 4.933 & 4.932  \\ %%% 4.93232  & 4.93319 & 4.93246 
\hline $128^3$ & 5.056 & 5.058 & --- \\%%5.05638 & 5.057525 & --- \\
\hline $256^3$ & 5.125 & 5.124  &  --- \\%%5.12479 & 5.12422  &  --- 
\hline  &  &  & \\
\end{tabular}
\end{centering}
\end{table}

\begin{table} 
\begin{centering}
\caption{ Like Table \ref{tab:photoncell}, but for $\rF / r_{\rm in}$. } \label{tab:packetcell}
\begin{tabular}{ |c |c |c |c |}
 \hline {\# Cells} &  {Low: $\langle r \rangle_F / r_{\rm in}$ } & {Medium: $\langle r \rangle_F / r_{\rm in}$ } &  {High: $\langle r \rangle_F / r_{\rm in}$ }\\
\hline $32^3$& 1.67 & 1.67&  1.67 \\ %{ \hline $32^3$& 1.67137 & 1.67171 &  1.67167 \\}
\hline $64^3$ & 1.63 & 1.63 & 1.63\\%{\hline $64^3$ & 1.63153  & 1.63149 & 1.63157 \\}
\hline $128^3$ & 1.61 & 1.61 & --- \\%{ \hline $128^3$ & 1.61179 & 1.61173 & --- \\ }
\hline $256^3$ & 1.60 & 1.60 &  --- \\%{\hline $256^3$ & 1.60179 & 1.60180 &  --- \\}
\hline  &  &  & \\
\end{tabular}
\end{centering}
\end{table}

%%%%%{\phjeditcom \noindent (8.) SCALING/COMPUTATION EXPLANATION }
Second, the dust density distribution must be sufficiently resolve the physical problem.   Here, convergence is not as rapid.  Varying the grid resolution (Table \ref{summarytable}) we observe significant variation in the integral quantities, with only a slow convergence {\cdmnew (error $\propto$(resolution)$^{-0.73}$)} that is, reassuringly, toward the DUSTY result. {
\cdmnew
%The slow convergence is not surprising;  the number of cells in our region scales with the cube of the linear resolution, and in turn 
As the number of photon packets must increase if the average number of packets per cell is to be maintained as a constant,}  
{\phjedittext %This shows 
highly accurate MC force evaluations can be quite costly.}
%%%%%5%%% DELETE THE BELOW SENTENCE %%%%%%%
%%%%Considering that the computational time scales {\ as the fourth power of the linear resolution times the number of photon packets per cell,} this implies a high computational cost to achieve accuracy of $\sim 10\%$.   
However, as we discuss below, we consider Hyperion's under-estimation of the radiation force to be a predictable systematic effect. 

\begin{table} 
\begin{centering}
\caption{A convergence study of Hyperion with grid resolution, and a comparison with  DUSTY \phjedit{ (for a flux conservation accuracy parameter of 0.001 and a maximum $\Delta \tau / \tau_{\rm max}$ of 0.025)}, for a physical problem created to match that in Tables \ref{tab:fluxcon} and \ref{tab:delTAUsc}.  Note that  $\Tdin$ converges toward the desired  value (1500\,K)  as $\Phi$ and $\rF$ converge. } \label{summarytable}
\begin{tabular}{ |c |c |c |c  |}
\hline {Resolution} &  { $\Phi$}  &  $\rF$   & {$\Tdin$ } \\
\hline $32^3$  & 4.73 & 1.67  &1209  \\ %{ \hline $32^3$  & 4.73383  & 1.67171  &1208.87195  \\ }
\hline $64^3$  & 4.93 & 1.63 &  1296\\ %{ \hline $64^3$  & 4.93319 & 1.63149 &  1295.6366\\ }
\hline $128^3$  & 5.06 & 1.61 & 1372 \\ %{\hline $128^3$  & 5.05725 & 1.61182 & 1372.45462   }
\hline $256^3$  & 5.12 & 1.60 &  1429 \\ %{ \hline $256^3$  & 5.12422  & 1.66018 ERROR --> 1.6017996432600392 &  1428.52965}
\hline DUSTY  & 5.20  & 1.59 & 1500 \\ %{ \hline DUSTY  & 5.20178  & 1.59441 & 1500.00000}
 \\
\hline
\end{tabular}
\end{centering}
\end{table}
%%%%%%%%%%%%%%% END OF TABLE
%%%NOTEBOOK PAGE 473 includes the information that we need now fo rthe scaling with respect to photons for cell.  

\subsubsection{Criteria for Monte Carlo convergence} \label{SS:MonteCarloConvergence}

Spatial resolution affects Monte Carlo radiation transfer codes like Hyperion in two distinct ways.  One is familiar to all numerical simulations: a physical problem is defined by only a few numbers per grid cell, so the problem itself (e.g., the spherical power-law dust density profile) converges to its ideal form only in the limit of infinite resolution.   Hyperion, for instance, treats the dust as uniform within each cell rather than enforcing our model $\rho\propto r^{-k}$.  

A second source of error arises from lack of resolution of the photon mean free path.  This can sometimes be handled in the diffusion approximation, but our problem is defined by a zone of starlight absorption in which dust reaches the maximum temperature $\Tdin$, and the peak wavelength of emission from this layer sets the Rosseland optical depth of the envelope.   This is most relevant in the optically thick regime, which is also where starlight is absorbed very close to $\rin$.  If the deposition of starlight energy is diluted by a lack of resolution, $\Tdin$ will be systematically underestimated.  The consequence is an artificial lowering {\cdm of the} Rosseland opacity $\kappa_R(\Tdin)$, as observed in Table \ref{summarytable}.  The radiation force $\Phi$ should be underestimated by the same factor, as in equation (\ref{eq:PhiIR}), even if the rest of the radiation transfer problem is handled perfectly. 

\begin{figure}
\begin{centering}
\includegraphics[width=0.50\textwidth]{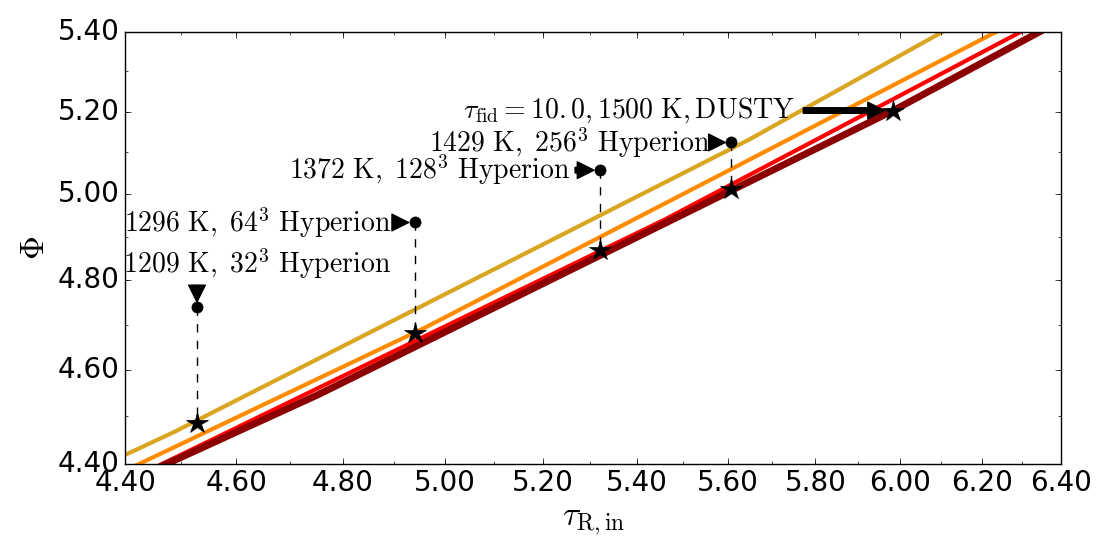}%{Comparison_Plot_JMA_revised_vertical.png}
\caption{
\cdmnew
An investigation {\phjedittext of the} force convergence in Hyperion Monte Carlo simulations of varying resolution (black circles).  In MC runs, $\Phi$ approaches an asymptote (top-right black star), identified using fully-resolved DUSTY simulations with the same dust column and inner temperature (1500\,K). Lower resolution MC runs underestimate $\Phi$.  This is partly because the inner temperature { \phjedittext is} underestimated when the mean free path of starlight is not resolved, as we demonstrate by connecting each run with a DUSTY simulation (black stars) of the same $\Tdin$ (dashed lines) and total dust column.   Yellow, orange, red, and dark red lines show the locus of DUSTY solutions for 1200, 1300, 1400, and 1500\,K, respectively.   
} \label{key_figure}
\end{centering}
\end{figure}

{\phjedittext We illustrate the resulting systematic underestimation of Monte Carlo forces and $\Tdin$ in Figure \ref{key_figure} for the case of $\taufid = 10$ and $\Tdin=1500$\,K, comparing against our ``ground truth" solution provided by DUSTY. 
%, which itself has been checked for reasonableness in comparison to the analytical estimates in \ref{doublephi}.  
Such an underestimation may be of particular relevance to radiative hydrodynamical simulations, where the characteristic scales of the hydrodynamical processes and the mean free path for the radiation often differ significantly; in such cases, additional resolution is required to properly capture the radiative transfer forces.   

The figure also {\cdmnew compares each resolution with well resolved DUSTY runs in which $\Tdin$ is adjusted to agree with each Hyperion run (as opposed to the 1500\,K target value).  We see that the tendency for Hyperion to underestimate $\Tdin$ at coarse linear resolution explains much of the discrepancy in $\Phi$.     }
%We see that the Hyperion temperature underestimates in the case of resolution produce $\Phi$ values intermediate between the DUSTY run with that same temperature and the ideal solution. Initially, this tends much closer to the DUSTY value for the underestimated temperature, but we observe convergence as the resolution improves.     
} 

%%%%% THIS IS ORIGNIAL TEXT BEING COMMENTED OUT
%%%%If correct, the effect of discretization should be much less noticeable at lower optical depths for which the starlight mean free path is not so short.   We tested this out with Hyperion runs of varying resolution and with $\taufid$ as low as {{$10^{-3}$.  {
%%%%%As expected, the discrepancy in $\Phi$ depends strongly on optical depth: for $\log_{10} \taufid = (-3,-2,-1,0,1)$ we find that the $32^3$ results underestimate DUSTY's $\Phi$ by $(0.57\%, 1.6\%, 0.7\%, 2.8\%, 8.7\%)$ whereas the $256^3$ results are $(0.11\%, 0.12\%, 0.23\%, 0.87\%, 1.2\%)$ low. } } }

%%%%Resolution affects Hyperion's determination of $\rF$ as well, but not as strongly:  $\rF$ decreases by {  4.2\% } going from $32^3$ to $256^3$ in the model with $\taufid=10$.  This undoubtedly reflects the fact that $\rF$ approaches a unique asymptotic value in each of the three regimes discussed in \S\,\ref{S:Analytics}, so it should be relatively insensitive to errors.

%%%%% AN ALTERNATE TAKE ON THESE PARAGRAPHS, BEING WRITTEN IN THE COMMENTS

\begin{figure}
\begin{centering}
\includegraphics[width=0.50\textwidth]{{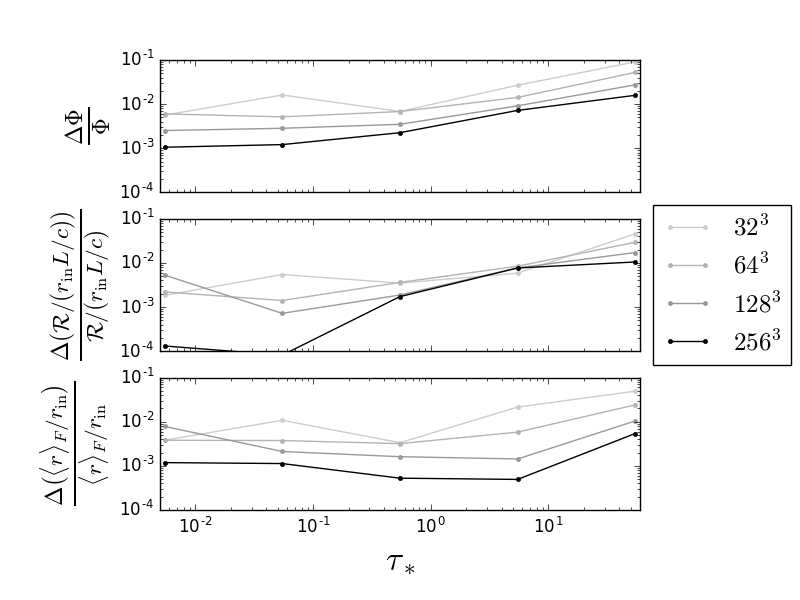}}%{fraction_plots}
\caption{ {\phjedittext The magnitude of fractional difference in the values of $\Phi$, $\langle r \rangle_F/r_{\rm in}$, and $\mathcal{R}/\left(r_{\rm in} L/c \right)$ as found with the Monte Carlo code Hyperion at varying resolutions, versus DUSTY models with a required flux accuracy parameter of 0.001.   }   } \label{fraction_figure}
\end{centering}
\end{figure}

{\phjedittext   This suggests that the effect of discretization should be much less noticeable at lower optical depths for which the starlight mean free path is not so short.  We tested this out with Hyperion runs of varying resolution and with $\taufid$ as low as $10^{-3}$. As expected, the discrepancy in each parameter depends strongly on optical depth.   For $\Phi$, we find that for $\log_{10} \taufid = (-3,-2,-1,0,1)$ that the $32^3$ results underestimate DUSTY's  $\Phi$  by  $(0.57\%, 1.6\%, 0.7\%, 2.8\%, 8.7\%)$ whereas the $256^3$ results are $(0.11\%, 0.12\%, 0.23\%, 0.87\%, 1.2\%)$ low, assuming a required flux accuracy of $0.001$.  

Resolution affects Hyperion's determination of $\rF$ as well, but not as strongly:  $\rF$ decreases by {  4.2\% } going from $32^3$ to $256^3$ in the model with $\taufid=10$.  This undoubtedly reflects the fact that $\rF$ approaches a unique asymptotic value in each of the three regimes discussed in \S\,\ref{S:Analytics}, so it should be relatively insensitive to errors.   

Finally, as noted in Equation \ref{virial-R}, for our problem, $\mathcal{R} = \Phi \frac{L}{c} \langle r \rangle_F$.  Therefore, it is unsurprising that in the optically thick regime, the errors $\mathcal{R}/\left(r_{\rm in} L/c \right)$ track in a similar manner to those of $\Phi$.   

The variation of these errors with optical depth is visualized in Figure \ref{fraction_figure}.  The radiative fluxes at the in radii $r_{\rm in}$ for these models are listed in Table \ref{tab:flux_tab}.
}

\begin{table}
\begin{centering}
\caption{Fluxes at $r_{\rm in}$ for $T_{\rm in} = 1500 \ {\rm k}$}
\label{tab:flux_tab}
\begin{tabular}{|c |c |}
\hline {$\taufid$} & {Flux in $10^6$ erg/cm$^2$/s}  \\
\hline $10^{-3}$ & 252 \\
\hline $10^{-2}$ & 246 \\
\hline $10^{-1}$ & 197 \\
\hline $10^{0}$ &  110 \\
\hline $10^{1}$ &  66.5\\
\end{tabular}
\end{centering}
\end{table}

%%% PROBABLY SHOULD TRUNCATE VALUES TO SOMETHING LIKE TWO SIGNIFICANT FIGURES

\subsection{Parameter space survey} \label{SS:ParamSpaceSurvey}

\begin{figure}
  \centering
  \subfloat[$\Phi$ vs.\ starlight optical depth]{\includegraphics[width=0.4\textwidth]{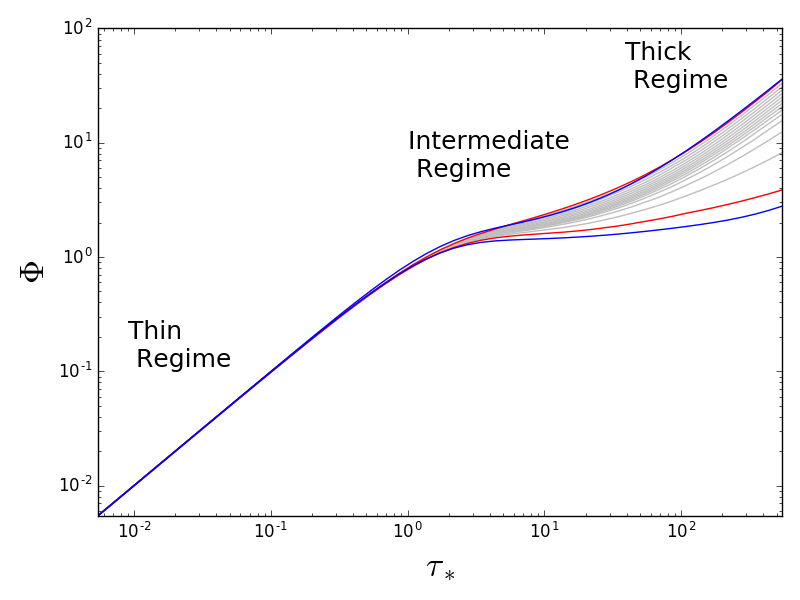}\label{fig:f1}}%%{phicomp_JMA_revised.png}
  \hfill
  \subfloat[$\Phi$ vs. \ { optical depth $\tau_{R,\rm in}$ calculated from the inner Rosseland opacity $\kappa_R(\Tdin)$}]{\includegraphics[width=0.4\textwidth]{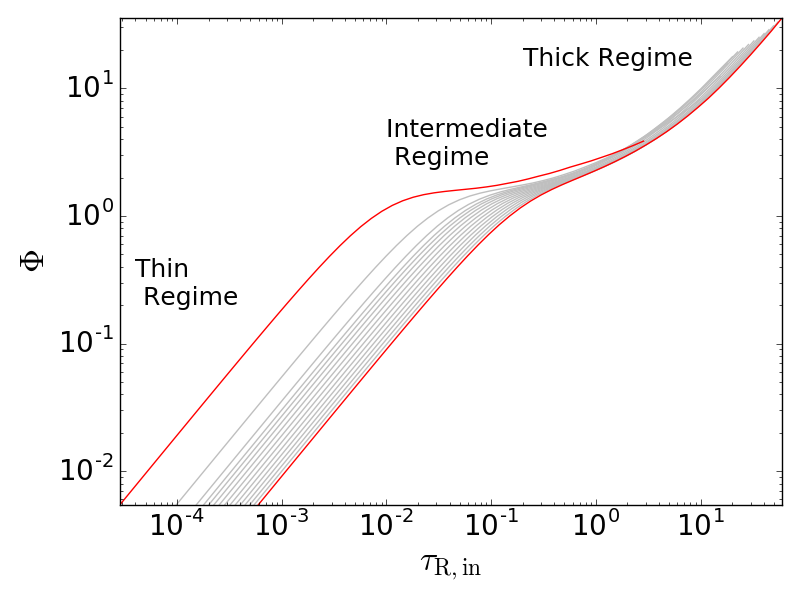}\label{fig:f2}} % {phi_R_compare_no_pred_JMA_revised.png}
  \caption{{{(a): Plot of the $\Phi_{\rm rad}$ parameter for radiation pressure force against the optical depth to starlight, $\tau_{\rm *}$, for 765 DUSTY models (with 15 different inner dust temperatures each at 51 different maximum optical depths), compared against analytical approximations for $\Phi_{\rm rad}$ utilizing the direct, scattering, and diffusion regimes. The corresponding values of $\tau_{\rm fid}$ range from 0.001 to 100.   The inner dust temperatures range from 100 K to 1500 K, at an interval of 100 K.  All models assume a geometry with $\rout =4\rin$ and $k = 1.5$.  The 100 K (lower) and 1500 K (upper) contours are highlighted in red, with the intermediate contours shown in gray.  Also shown in blue is an analytical approximation for  $\Phi_{\rm rad}$ for 100 K and 1500 K models.  We see the convergence of the analytical approximation and the DUSTY solution in the direct radiation and the diffusion limits.   In the intermediate scattering regime, the analytical approximation underestimates DUSTY.   (b):  The same $\Phi_{\rm rad}$ contours, but now plotted against the axis of  $\tau_{\rm R,in}$, the Rosseland mean optical depth of the dust to radiation reprocessed at the inner temperature.  Once again, the 100 K (leftmost) and the 1500 K (rightmost) contours of the models are highlighted in red.    } } }  \label{doublephi}
\end{figure}

\begin{figure} 
\centering
{\includegraphics[width=0.5\textwidth]{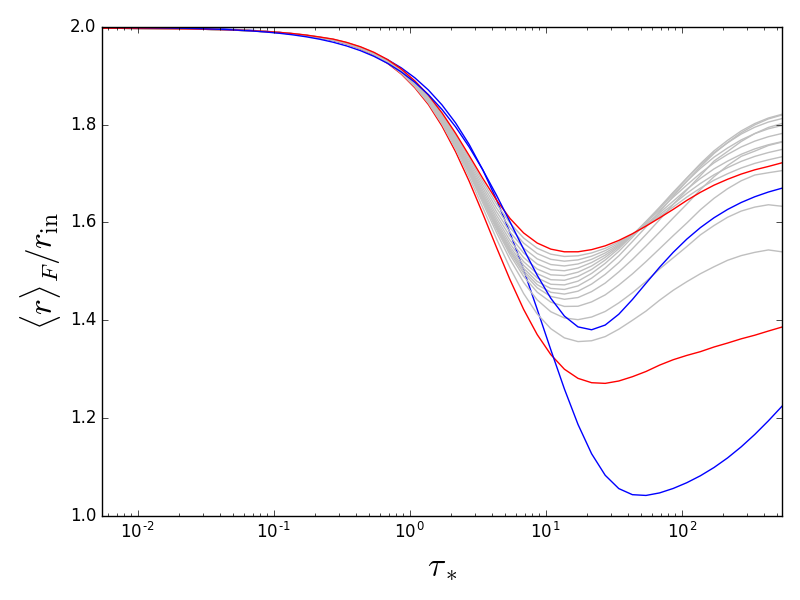}\label{fig:f1}}
\hfill
\caption{ {  Plot of the $ \langle r \rangle_F $ parameter for radiation pressure force against the optical depth to starlight, $\tau_{\rm *}$, for 765 DUSTY models (15 different inner dust temperatures each at 51 different maximum optical depths), compared against analytical approximations across the direct, scattering, and diffusion regimes. The corresponding values of $\tau_{\rm fid}$ range from 0.001 to 100.   The inner dust temperatures ranges over an interval from 100 K to 1500 K with 100 K increments.  All models assume a geometry of $n =4$ and $k = 1.5$.  The 100 K (lower) and 1500 K (upper) contours are highlighted in red, with the intermediate contours shown in gray.   Also shown in blue is an analytical approximation for  $\langle r \rangle_F /r_{\rm in}$ for 100 K and 1500 K models.} 
  %(b.)  The same  $ \langle r \rangle_F $ contours, but now plotted against the axis of $\tau_{\rm R, in}$, as in Figure \ref{doublephi}(b).   
} 
\label{doubley}  
\end{figure}

With the scaling behaviors examined, we now examine the overall  behaviour of the radiation transfer solutions. %%% Behaviour - is this a US/CAN English difference?      
Figures \ref{doublephi} and  \ref{doubley} present the variation of $\Phi$  and $\rF$, respectively, across the range  $10^{-3}  <\taufid<10^2$ and 
%several decades of maximum dust envelope optical depths (0.001 to 100.0 = $\tau_{\rm fid}$) for a range of 15 temperatures inner dust temperatures ranging from 
100\,K$<\Tdin<$1500\,K. %Figure \ref{doubley} does the same, except for $\langle r \rangle _F$.
{  We plot DUSTY results { (with the adopted parameters) }  against { $\tau_*$ } in panel (a) of Figure \ref{doublephi}, and against {  $\tau_R(\Tdin) = [\kappa_R(\Tdin)/\kappa_{\rm *}] \tau_*$ } in panel (b);  the  two plots are meant to illustrate that $\Phi$ depends on $\tau_*=5.47\taufid$ in the optically thin regime}  %% WE COULD just recast it to plotted against taustar to make the relation more transparent, which is what my original plots already did.   
 and on $\tau_R(\Tdin)$ in the thick regime. { Figure \ref{doubley} makes a similar point.}  These dependences are hard-wired into the analytical predictions, equations (\ref{eq:combined-Phi}) and (\ref{eq:combined-rF}), which we overplot in each figure. { (To avoid clutter we plot these predictions only for $\Tdin=100$\,K and 1500\,K.) } 

These figures make very clear that the three radiation transfer regimes discussed in \S\,\ref{S:Analytics} apply to the real problem as well as the idealized one.  As for our analytical predictions: { while for low temperatures the error in $\Phi$ is large, with a peak of $\approx 50$\% for 200 K at $\tau_* =274$ over the 100 K and 200 K range, there is also a peak error of $\approx 30\%$ at $\tau_* =547$ at a temperature of 500 K on the range of 300 K to 800 K,  and a peak error of 8.5\% also at $\tau_* =547$ at a temperature of 900 K over the range of 900 K -1500 K. }

%%% OLD y_eff values:   at tau = 0.001, 0.0109 for DUSTY yeff
%% Hyperion $\approx 0.0056$  --> 1.025 = 0.0056/0.005466

\section{Conclusions } \label{S:Conclusions}

 We start with our key findings. 
 First, for a spherical dust envelope the overall importance of radiation force is captured by a couple  integral quantities: a normalized radial force $\Phi$ (formally equivalent to the net flux-averaged optical depth) and a force-averaged radius $\rF$.  These combine to give the radiation term in the virial theorem, ${\cal R} \equiv\int {\mathbf r}\cdot d{\mathbf F} =\Phi \rF L/c$, and so they directly affect the dynamical evolution when radiation forces matter at all.   We stress this point because radiation forces are frequently described only in terms of a net force.  Assuming the force is applied on the largest scales gives an overestimate for $\cal R$, because $\rF$ tends to be of order the innermost radius, or (in the optically thin case) intermediate between inner and outer radii.  %% One comment thought is that since most of my models are rout/rin = 4, we may not have had as thorough an exploration of the deposition as we might have liked 

Second, the difference in opacity between starlight and thermal infrared radiation opens an intermediate regime in which starlight is easily scattered and absorbed, but thermal emission escapes.  For this reason,
the radiation transfer problem breaks into thin, intermediate, and thick regimes.  Each of these can be understood analytically in its asymptotic form, and we provide approximate formulae for $\Phi$, $\rF$, and $\cal R$ that combine these forms into a single expression {\phjedit valid to a few tens of percent (and within 10\% at higher temperatures).} 

{
Third, we 
{\cdmnew calibrate the accuracy with which force parameters are determined in three-dimensional Monte Carlo simulations of moderate spatial resolution, by comparing results from the Hyperion MC code (run in Cartesian geometry) against highly resolved calculations by DUSTY (run in spherical symmetry).  
%compare the DUSTY radiation transfer code, which is self-adaptive and optimized for our spherical problem, against the Hyperion Monte Carlo code. 
Although MC forces converge toward the physical solution, 
%both codes converge toward a physical solution,
this comparison reveals 
that Monte Carlo underestimates radial forces when the starlight mean free path is not well resolved.  Part of this error comes from the fact that stellar luminosity is then deposited in too thick a layer, leading to an underestimate of the maximum dust temperature. Because of the wavelength dependence of dust opacity, the net result is an underestimate of the net radiation force. }

%
%%%%%{\phjcommentnew - This next paragraph pertains to their complaint about how we dealt with computational resources}
%

This may generate a tension in the design of Monte Carlo simulations, between the desire to resolve the starlight mean free path and the need to allocate numerical resources.   
{\phjedittext This is particularly relevant to hydrodynamic models for which resolving the mean free path of starlight would be prohibitive.}
%Although integral quantities like $\Phi$ and $\rF$ depend very weakly on the number of  photon packets propagated, and although Monte Carlo codes parallelize efficiently with respect to the number of such packets, the total computational cost scales as the fourth power of grid resolution if the number of packets per cell is held constant.  Furthermore, this parallelization requires a copy of the density grid to exist on each node, limiting the number of cells that can be defined.  
One solution would be to locally refine the grid using the starlight mean free path as a refinement criterion.   Another would be to implement a sub-grid model for the dust temperature profile.  A third would be to apply a correction factor to remove the systematic effect of poor resolution on the radiation forces.  

Although we have only considered spherically symmetric dust profiles, we can comment on non-spherical effects.  Clearly, segregating dust into clumps and opening paths of lower optical depth will reduce the trapping of radiation, lowering the net radiation force and $\Phi$, especially in the optically thick regime.  On the other hand, the same effects tend to increase the radial scale $\rF$ on which radiation forces are applied, potentially by a large factor.  The net effect on $\cal R$ is not immediately obvious.   We intend to return to these questions in a future publication on non-spherical and inhomogeneous density distributions. 

{~}

\noindent This work was supported by a Connaught fellowship (PJ) and an NSERC Discovery Grant (CDM), and was enabled in part by support provided by Compute Ontario and Compute Canada for calculations done on the SciNet facility in Toronto.  We thank our referee for comments that helped us improve this work.   

\appendix 

\section{Scattered light} \label{S:ScatteredLightAppendix}
%\subsection{}

Here we consider the scattered starlight.  Because in our numerical investigations we implement isotropic scattering, the scattered light is close to isotropic at all radii.  It can therefore be treated with Eddington's approximation, in which the specific intensity in direction $\hat n$ is a linear function of $\mu = \hat n\cdot \hat r$ at every radius.   The equations are exactly the same as those presented by  \citeauthor{1986rpa..book.....R}, with the exception that the direct illumination by starlight adds a new source of scattered radiation.  In this section, for additional clarity, we rename the total optical depth of the dust envelope $\tau_{\rm \nu,\max}$ and the local optical depth from the centre  as $\tau_\nu$.

Defining $\tauhatnu = (3 \epsilon_\nu )^{1/2} \tau_\nu$, where $(1-\epsilon_\nu) = a_\nu$ is the albedo, the mean scattered intensity $J_\nu$ satisfies
\begin{equation} \label{eq:ScatteringDiffusion} 
J''_\nu -  J_\nu + \left( \Sdirhatnu +B_\nu\right)=0
\end{equation} 
where prime denotes $d/d\tauhatnu$, $B_\nu$ is the thermal radiation at the local dust temperature, and 
\begin{equation}\label{eq:ScatteringSource} 
\Sdirhatnu = {1-\epsilon_\nu\over \epsilon_\nu} {L_{*,\nu}\over (4 \pi)^2 r^2 } e^{-\tau_\nu}
\end{equation}
is the additional source term. 
With the inner boundary condition $J'(0)=0$ corresponding to no net scattered flux at the origin (we take $\tauhatnu$ increasing outward), equation 
(\ref{eq:ScatteringDiffusion})  has the explicit solution 
\begin{equation} \label{eq:ScatteringDiffusionGenSoln} 
J_\nu = C_\nu \cosh(\tauhatnu)  + \int_0^\tauhatnu \left[\Sdirhatnu(\tau') + B_\nu(\tau') \right]\sinh({\tauhatnu-\tau'})\, d\tau'. 
\end{equation}
The integration constant $C_\nu$ is determined by the condition of zero incoming flux at the outer boundary.  In the two-stream approximation as described by \citeauthor{1986rpa..book.....R} this condition is $3^{1/2} J_\nu +  dJ_\nu/d\tau_\nu =0$, which corresponds to $J_\nu + \epsilon_\nu^{1/2}J_\nu'=0$, at the maximum effective optical depth $\tauhatmax$.  This implies
%\begin{widefig}[.1in]
\begin{eqnarray} \label{eq:ScatteringIntegrationConstant}
\centerline{$
C_\nu =   { \int_0^\tauhatmax \left[ \sinh(\tauhatmax-\tau') +\epsilon_\nu^{1/2}\cosh(\tauhatmax-\tau') \right]\,  \left[\Sdirhatnu(\tau')+ B_\nu(\tau') \right] \,d\tau'    
          \over \cosh(\tauhatmax) + \epsilon_\nu^{1/2}\sinh(\tauhatmax)}. 
         $ }\nonumber\\
\end{eqnarray}

%\end{widefig}

Because the star is much hotter than the dust, and we are concerned here with the peak frequencies for scattered starlight, we neglect $B_\nu$ in practice.   This has the benefit that equations (\ref{eq:ScatteringDiffusionGenSoln}) and (\ref{eq:ScatteringIntegrationConstant}) can be evaluated directly, without solving self-consistently for the dust temperature distribution.

Our goals involve the radiation force due to scattered starlight.  For this we require the scattered flux, which in Eddington's approximation is given by $\Fscatnu = -(4\pi/3)dJ_\nu/d\tau_\nu = -4\pi (\epsilon_\nu/3)^{1/2} J'_\nu$, where
 \[J'_\nu(\tauhatnu)=C_\nu \sinh(\tauhatnu) - \int_0^\tauhatnu \left[\Sdirhatnu(\tau')+B_\nu(\tau')\right] \cosh(\tauhatnu-\tau')\,d\tau'.\] 
The luminosity of this radiation is $\Lscatnu = 4\pi r^2 \Fscatnu$, and its differential force is $d\Forcescatnu = (\Lscatnu/c) d\tau_\nu $.    Its normalized total force at frequency $\nu$ is then $\phiscnu = \Forcescatnu/(L_{*\nu}/c)$, and its force-averaged radius is $\rbarscatnu = (\int r\,d\Forcescatnu)/\Forcescatnu$.    Another ingredient is the relation between $r$ and $\tau$, for which the simple cloud model $\rho = \rhoin (\rin/r)^k$ implies $ r = \rin \left(1 - {\tau_\nu/\tauinfnu} \right)^{-\frac{1}{k-1}} $ where $\tauinfnu = \rhoin \rin \kappa_\nu/(k-1)$.  Finally, we have two equivalent expressions for the maximum optical depth: 
\[ 
\tau_{\nu,{\rm max}} ~\simeq~ \left[ 1 - \left(\rin\over \rout \right)^{k-1} \right]\tauinfnu  ~=~{\kappa_\nu\over \kappa_{\rm fid}}  \tau_{\rm fid}. 
\] 
%(It may be more correct to subtract 2/3 from these values, as diffusion occurs only within the photosphere, but applying this offset neglects radiation force in the optically thin region.) 

The outcome of this analysis is plotted in Figure \ref{fig:EddingtonScattering} for our fiducial case ($k=1.5, \rout = 4 \rin$).  We see that the force is increased by about a factor of two over the direct force of starlight for reasonably high albedos, and that the force is applied at a location that is at most a couple times the inner boundary, decreasing toward $\rin$ as the optical depth increases.  Both of these trends are markedly different from the diffusive force due to dust emission. 
 
\begin{figure}
%\vskip -0.2cm
\centering
$\begin{array}{cc}
\includegraphics[angle=0,width=8.5cm]{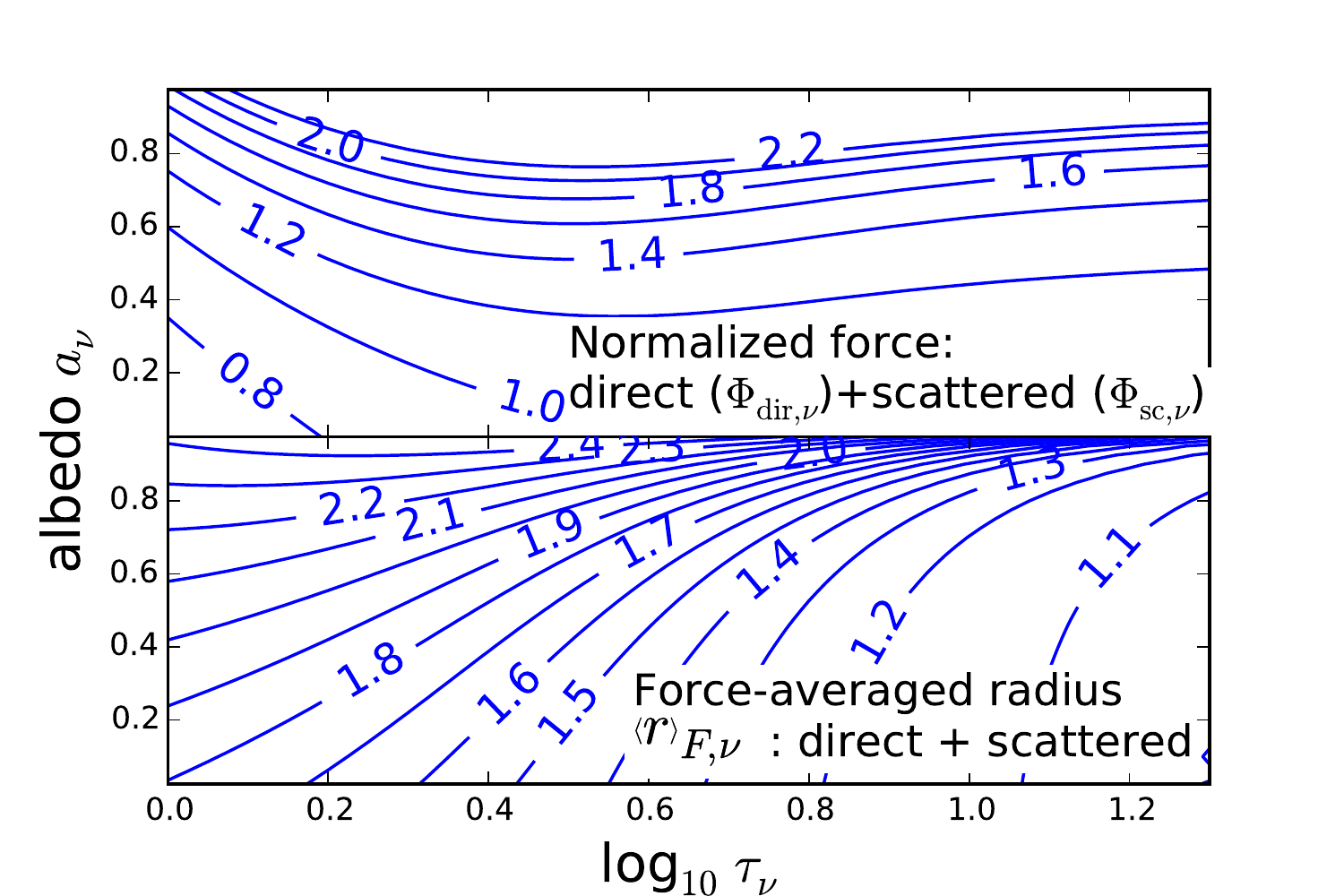} 
\end{array}$
\caption{Normalized force (top) and force-averaged radius (bottom) for direct and scattered starlight, using Eddington's approximation for the scattering, for the case $k=1.5, \rout = 4\rin$.  }
\label{fig:EddingtonScattering}
\vspace{-0.5cm}
\end{figure}

\subsection{Limit of high optical depths}  \label{SS:HighTauScattering}
The case of a very optically thick cloud (to starlight) is of particular interest, as scattering is most important in this regime.   So long as the cloud is also {\em effectively} optically thick, so that $\tauhatmax \gg 1$,  the force due to scattered light becomes analytical, because all the light is absorbed near the inner boundary and one can take $r=\rin$, independent of $\tau_\nu$.   Then $\Sdirhatnu = S_0 e^{-\tau_\nu}$ where $S_0 = (\epsilon_\nu^{-1}-1)L_{*,\nu}/(4\pi \rin)^2$.    The requirement $J'\rightarrow 0$ as $\tauhatnu\rightarrow \infty$ implies $C_\nu=\int_0^\infty S(\tauhatnu) e^{-\tauhatnu} d\tauhatnu = S_0/[1 + ({3\epsilon_\nu})^{-1/2}]$, so $J' = -S_0 (3\epsilon_\nu)^{1/2}(e^{-\tau_\nu} - e^{-\tauhatnu})/(1-3\epsilon_\nu)$.   Computing the force and comparing to the photon momentum, the ratio is 
\begin{equation} \label{eq:ForceAtHighOpticalDepths}
\phiscnu = {1-\epsilon_\nu \over 1 + (3\epsilon_\nu)^{1/2} } 
\end{equation} 
and in this limit, all of the starlight momentum is transferred to the cloud ($\phidirnu=1$).   Equation (\ref{eq:ForceAtHighOpticalDepths}) shows that when scattered light is absorbed near the inner boundary, its force is comparable to that of the direct radiation.  (For the net force of scattered radiation to become significantly larger, the scattered photons must penetrate beyond the inner radius.)

{ We note that, at the inner boundary, the mean intensity of scattered radiation is $(1-\epsilon_\nu)/(\epsilon_\nu + \sqrt{\epsilon_\nu/3})$ times that of the direct starlight in this limit.  Backscatter should therefore push the sublimation radius outward by the factor $\left[ (1+\sqrt{3/\epsilon_*}) / (1 + 1/\sqrt{3\epsilon_*} ) \right]^{1/2}$.} 

{
\section{Diffusion of thermal infrared light} \label{S:Diffusion} 

At the risk of rehashing familiar material (e.g., \citealt{km09} eq.~33), we consider $\tau_R(\Tdin)\gg 1$ and work in the diffusion approximation, $d\Prad = - L\,d\tau_R/(4\pi r^2 c)$ where $d\tau_R = \kappa_R \rho \, dr$ for Rosseland opacity $\kappa_R(T)$.   Integration yields the profile of temperature and radiation pressure $\Prad = a T^4/3$:  
\begin{equation}\label{eq:DiffusionSoln_general}  
\int_{\Prad(\rph)}^{\Prad(r)} {d\Prad\over \kappa_R} = {L\over 4\pi c} \int_{1/\rph}^{1/r} \rho\, d(r^{-1}). 
\end{equation} 
 The effective photosphere $\rph$ is the radius from which the Rosseland optical depth to infinity is roughly unity and the temperature is set by the Stefan-Boltzmann relation $L\simeq 4\pi \rph^2 \sigma_{\rm SB} T(\rph)^4$ so that $\Prad(\rph) \simeq L/(3\pi \rph^2 c)$.  Note that, for high enough optical depth, $\rph$ approaches the outer boundary $\rout$ if one exists.   If we adopt an opacity power law $\kappa_R(T) \propto T^\beta$ (in addition to the density power law $\rho(r)\propto r^{-k}$) then, integrating equation (\ref{eq:DiffusionSoln_general}),
\begin{equation} \label{eq:DiffusionSoln_complicated} 
{\Prad \over \kappa_R}  =  {1-\beta/4 \over 1 + k} {L \rho\over 4 \pi c r} \left[1- \left(r\over \rph\right)^{k+1} \right] + {L\over 3\pi \rph^2 \kappa_R(\rph) c}. 
% + {\Prad(\rph) \over \kappa_R(\rph) }. 
\end{equation} 
The last term is of order $(\rph/r)^{-2} (\kappa \rho r)^{-1}$ relative to everything else: it can safely be ignored in regions of high optical depth.   Likewise the second term in brackets is negligible for $(r/\rph)^{k+1}\ll1$, so it can often neglected near the inner boundary. One is then left with the inner power law profile 
\[
{\Prad\over \kappa_R} \simeq {(1-\beta/4)L\rho \over 4\pi (1+k) c r}
\]
 in which $T\propto r^{-(k+1)/(4-\beta)}$.

The force distribution is particularly simple  as $dF_{\rm IR} =(L/c) d\tau_R$ in the diffusion approximation, so $\PhiIR = \tau_R$ modulo a small offset arising from the photosphere and optically thin region.   The inner power law solution suffices to estimate $\PhiIR$, because the force is concentrated in the densest, hottest regions near the inner boundary; this gives equation (\ref{eq:PhiIR}). 

The force-averaged radius $\rF$ is also determined by central conditions, although not to the same degree as $\PhiIR$:  $\rF$ takes the optically thick limit 
\begin{eqnarray} \label{eq:rF_diffusive}
 \rFthick &=& { \int_\rin^\rph \left[ 1 - (r/\rph)^{k+1}\right]^{\beta\over 4-\beta} r^{1-\left[{\beta(k+1)\over 4-\beta}+k \right]}\,dr \over 
 \int_\rin^\rph \left[ 1 - (r/\rph)^{k+1}\right]^{\beta\over 4-\beta} r^{ -\left[{\beta(k+1)\over 4-\beta}+k \right]}\,dr } ~~~\\
&~ &
 \xrightarrow[]{\,\rph\gg\rin \,} 
 2{ 2(k-1)+\beta \over  4-\beta} \rin.  \nonumber 
\end{eqnarray} 
The second expression is valid only insofar as the inner power law solution holds to radii well beyond $\rF$, and so should not be used in our fiducial problem. }

%\eject
\vskip 1.2in
\bibliographystyle{plainnat}
\bibliography{JumperMatzner-1.bib} 

\end{document}